\documentclass[12pt,a4paper,nofootinbib,prd]{revtex4}

\usepackage{amsmath}

\newcommand*{\Dz}{$D^0$\ }
\newcommand*{\Dzbar}{$\overline{D^0}$\ }
\newcommand*{\DDbar}{\Dz- \Dzbar}

\begin{document}

\title{Interpreting experimental bounds on \DDbar mixing\\*
  in the presence of CP violation}

\author{Guy \surname{Raz}}
\affiliation{Particle Physics Department \\ Weizmann Institute of
  science \\ Rehovot 76100, Israel}
\email{guy.raz@weizmann.ac.il}

\begin{abstract}
  We analyse the most recent experimental data regarding \DDbar
  mixing, allowing for CP violation. We focus on the dispersive part
  of the mixing amplitude, $M^D_{12}$, which is sensitive to new
  physics contributions. We obtain a constraint on the mixing
  amplitude: $\left|M^D_{12}\right| \leq 6.2\times 10^{-11}$ MeV
  at~$95\%$ C.L.~. This constraint is weaker by a factor of about
  three than the one which is obtained when no CP violation is
  assumed.
\end{abstract}

\maketitle


\section{Introduction}
\label{sec:introduction}

The ongoing searches for \DDbar mixing
\cite{Link:2000cu,Godang:1999yd,Link:2000kr,Schwartz:2000mm,
  Csorna:2001ww,Abe:2001ed,Aubert:2001aq,babar} have not yet detected
a signal of such mixing. Thus, the experimental data place an upper
bound on the mixing amplitude.  The value of this upper bound,
however, depends on the assumptions one makes when analysing the
experimental results. Specifically, the question of CP violation in
\DDbar mixing has an important impact on the final answer. Most often,
\DDbar mixing experiments are analysed assuming no CP violation.
While this assumption is valid for the standard model, it does not
hold for many new physics models.\footnote{In fact, since it has been
  recently suggested that the \DDbar mixing amplitude may be large
  even in the standard model \cite{Falk:2001hx}, CP violation may be
  the most valuable clue for new physics in this system.} (See, for
example, the supersymmetric models
in~\cite{Nir:1993mx,Leurer:1994gy}.)  Obviously, if the constraint on
\DDbar mixing is to be used to test such new physics models, the
experimental data should be interpreted in an appropriate
framework~\cite{Blaylock:1995ay}. We therefore present here the
analysis of experimental results allowing for CP violation in
mixing.\footnote{We do not consider, however, CP violation in \Dz
  decays, which is absent in most new physics extensions
  \cite{Bergmann:1999pm}.}

The organization of this work is as follows: In
section~\ref{sec:notation-formalism} we present our formalism. We
review the most recent experimental data in
section~\ref{sec:review-recent-ddbar} , and perform the analysis in
section \ref{sec:interpr-exper-data}. We conclude in
section~\ref{sec:conclusions}.

\section{Notation and formalism}
\label{sec:notation-formalism}

We follow mostly the formalism of ref.~\cite{Bergmann:2000id}.
The mass eigenstates are given by
\begin{equation}
 \label{eq:1}
  \left|D_{1,2}\right\rangle =  p\left|D^0\right\rangle \pm q\left|\overline{D^0}\right\rangle \;,
\end{equation}
The mass and the width differences are parameterized as follows:
\begin{equation}
 \label{eq:2}
 x\equiv \frac{m_2-m_1}{\Gamma}\;\;,\;\;\hfill y\equiv \frac{\Gamma_2-\Gamma_1}{2\Gamma}\;,
\end{equation}
with the average mass and width defined as 
\begin{equation}
 \label{eq:3}
 m\equiv \frac{m_1+m_2}{2}\;\;,\;\;\Gamma\equiv \frac{\Gamma_1+\Gamma_2}{2}\;.
\end{equation}
We define the \Dz and \Dzbar decay amplitudes by
\begin{equation}
 \label{eq:4}
 A_f \equiv \left\langle f \right| \mathcal{H}_d \left| D^0
 \right\rangle\;\;,\;\;%
 \bar{A}_f\equiv \left\langle f \right| \mathcal{H}_d \left| \overline{D^0}
 \right\rangle\;,
\end{equation}
and the complex observable $\lambda_f$ as
\begin{equation}
  \label{eq:5}
  \lambda_f\equiv\frac{q}{p}\frac{\bar{A}_f}{A_f}\;.
\end{equation}

In almost all models, CP violation in decay of the relevant modes can
be safely neglected \cite{Bergmann:2000id,Bergmann:1999pm} (for
an exception, see~\cite{D'Ambrosio:2001wg}), leading to
\begin{equation}
 \label{eq:6}
 A_f=\bar{A}_{\bar{f}}\;.
\end{equation}

Now we can parametrize the effects of indirect CP violation in the
relevant decay processes: The doubly-Cabibbo-suppressed (DCS)
\Dz$\longrightarrow K^+\pi^-$, the singly-Cabibbo-suppressed (SCS)
\Dz$\longrightarrow K^+K^-$, the Cabibbo-favoured (CF)
\Dz$\longrightarrow K^-\pi^+$, and the three conjugate processes. We
denote\footnote{Note that $A_m$ in our definition is twice smaller
  than the $A_m$ used by CLEO~\cite{Godang:1999yd}.}
\begin{eqnarray}
  \label{eq:7}
  {\left|q/p\right|}^2 &=& 1+2 A_m \;, \\
  \lambda^{-1}_{K^+ \pi^-} &=&
  \sqrt{R_D}\left(1-A_m\right)e^{-i(\delta+\phi)}\;, \\
  \lambda_{K^- \pi^+} &=&
  \sqrt{R_D}\left(1+A_m\right)e^{-i(\delta-\phi)}\;, \\
  \lambda_{K^+ K^-} &=&
  -\left(1+A_m\right)e^{i\phi}\;,
\end{eqnarray}
where the $\phi$ and $\delta$ are the weak phase and the strong phase,
respectively, and
\begin{equation}
 \label{eq:8}
 R_D = \left|\frac{A_{K^+\pi ^-}}{\bar{A}_{K^+\pi ^-}}\right|^2 =
 \left|\frac{\bar{A}_{K^-\pi ^+}}{A_{K^-\pi ^+}}\right|^2  \;.
\end{equation}

Next we define
\begin{eqnarray}
 \label{eq:xpyp}
   x'& \equiv & x \cos \delta + y \sin \delta \;, \\
   y'& \equiv & y \cos \delta - x \sin \delta \;. \nonumber
\end{eqnarray}

The rates of the DCS, SCS and CF decays 
are expanded for short times $t\lesssim 1/\Gamma$ as
\begin{multline}
  \label{eq:9}
  \Gamma [D^0(t) \longrightarrow K^+\pi^-] = e^{-\Gamma
    t}\left|A_{K^- \pi^+}\right|^2 \\*
  \times \left[R_D+\sqrt{R_D}(1+A_m)(y'\cos\phi 
    -x' \sin \phi )\Gamma t\right. 
  \left.+\frac{1+2A_m}{4}(y^2+x^2){(\Gamma
      t)}^2\right] \;,
\end{multline}
\begin{multline}
  \label{eq:10}
  \Gamma [\overline{D^0}(t) \longrightarrow K^-\pi^+] = e^{-\Gamma
    t}\left|A_{K^- \pi^+}\right|^2 \\*
  \times \left[R_D+\sqrt{R_D}(1-A_m)(y'\cos\phi 
    +x' \sin \phi )\Gamma t\right. 
  \left.+\frac{1-2A_m}{4}(y^2+x^2){(\Gamma
      t)}^2\right] \;,
\end{multline}
\begin{equation}
 \label{eq:11}
 \Gamma [D^0(t)\longrightarrow K^+K^-] = e^{-\Gamma
   t}\left|A_{K^+K^-}\right|^2 
 \times \left[1-(1+A_m)(y'\cos\phi - x' \sin
   \phi)\Gamma t\right] \;,
\end{equation}
\begin{equation}
 \label{eq:12}
 \Gamma\left[D^0(t)\longrightarrow K^-\pi^+\right] =
 \Gamma\left[\overline{D^0(t)}\longrightarrow K^+\pi^-\right] = 
 e^{-\Gamma t}\left|A_{K^-\pi^+}\right|^2 \;. 
\end{equation}

Several experiments measure the parameter $y_{CP}$, defined by
\begin{equation}
 \label{eq:13}
 y_{CP}=\frac{\tau(D^0\longrightarrow
 K^-\pi^+)}{\tau(D^0\longrightarrow K^+K^-)}-1 \;,
\end{equation}
with $\tau$ being the measured lifetime fitted to a pure exponential
decay rate for the specific modes~\cite{Link:2000cu,Bergmann:2000id}.
If CP is a good symmetry in the relevant processes, this definition of
$y_{CP}$ corresponds to
\begin{equation}
 \label{eq:14}
 y_{CP}\equiv\frac{\Gamma(\text{CP even})-\Gamma(\text{CP
 odd})}{\Gamma(\text{CP even})+\Gamma(\text{CP odd})}\;,
\end{equation}
since then the $K^+K^-$ state is an even CP state and the $K^-\pi^+$
state is an equal mixture of CP even and CP odd states. By fitting the
decay rates in~\eqref{eq:11} and~\eqref{eq:9} to exponents, and
expanding for small $A_m$ we get~\cite{Bergmann:2000id}:
\begin{equation}
 \label{eq:ycpth}
 y_{CP}=y \cos \phi - A_m\, x \sin \phi\;.
\end{equation}

We are interested in the dispersive part of the mixing amplitude,
$M^D_{12}$: Short distance contribution from new physics can affect
$M^D_{12}$ in a significant way. In terms of measurable quantities,
$\left|M^D_{12}\right|$ is given by~\cite{Branco:1999fs}
\begin{equation}
  \label{eq:15}
  \left|M^D_{12}\right|^2=\frac{4{(\Delta
  m)}^2+{A_m}^2{(\Delta
  \Gamma)}^2}{16(1-{A_m}^2)} \; ,
\end{equation}
or, using eq.~\eqref{eq:2},
\begin{equation}
  \label{eq:16}
  \left|M^D_{12}\right|^2=\Gamma^2\frac{x^2+A_m^2y^2}{4(1-A_m^2)}
  \; .
\end{equation}

\section{Experimental data on \DDbar\ mixing}
\label{sec:review-recent-ddbar}

The neutral $D$ system is studied by various experiments. First, the
CLEO experiment~\cite{Godang:1999yd} measures the rates~\eqref{eq:9},~\eqref{eq:10}:
\begin{eqnarray}
 \label{eq:cleo}
 R_D &=& (0.48\pm 0.13)\% \;,  \nonumber \\
 y'\cos\phi &=& (-2.5^{+1.4}_{-1.6})\% \;, \nonumber \\
 x' &=& (0.0\pm 1.5)\% \;,  \\
 2\,A_m &=& 0.23^{+0.63}_{-0.80} \;, \nonumber \\
 \sin\phi &=& 0.00 \pm 0.60 \;. \nonumber
\end{eqnarray}

The FOCUS experiment~\cite{Link:2000kr} provides a measurement of the
ratio between the branching ratio of the DCS and CF decays.  This
measurement is consistent with CLEO data at the level of $\sim
0.8\sigma$. However, as no direct measurement of the parameters is
done, no stronger bounds on the parameters result.

The value of $y_{CP}$ is measured by the various experiments.
Table~\ref{tab:y_cp} presents the various results.
\begin{table}
  \caption{Measurements of $y_{CP}$.}
  \label{tab:y_cp}
  \begin{ruledtabular}
    \begin{tabular}{l c}
      Experiment & Value\\
      \hline
      FOCUS~\cite{Link:2000cu} & $(3.42\pm 1.39 \pm 0.74)\%$ \\
      E791~\cite{Schwartz:2000mm} & $(0.8 \pm 2.9 \pm 1.0)\%$ \\
      CLEO\cite{Csorna:2001ww} & $(-1.2 \pm 2.5 \pm 1.4)\%$ \\
      BELLE~\cite{Abe:2001ed} & $(-0.5 \pm 1.0 ^{+0.7}_{-0.8})\%$ \\
      BABAR~\cite{babar} & $(1.4 \pm 1.0 ^{+0.6}_{-0.7})\%$ \\ 
    \end{tabular}
  \end{ruledtabular}
\end{table}
The world weighted average of $y_{CP}$ is hence:
\begin{equation}
  \label{eq:ycpexp}
  y_{CP}=(1.0 \pm 0.7)\% \;.
\end{equation}

\section{Interpretation of the experimental data}
\label{sec:interpr-exper-data}


Our aim is to constrain the \DDbar\ mixing amplitude $M^D_{12}$. First
we combine \eqref{eq:xpyp} and \eqref{eq:ycpth} to get
\begin{equation}
  \label{eq:17}
  y_{CP}+A_m\sin \phi \left(x' \cos \delta + y' \sin
  \delta\right)=y'\cos\phi\cos\delta-x'\cos\phi\sin\delta \ .
\end{equation}
The measured values of~\eqref{eq:cleo} and~\eqref{eq:ycpexp} can be
used to constrain $\cos\delta$. Assuming first\footnote{A similar
  procedure was followed in ref.~\cite{Bergmann:2000id}.} $A_m=0$ and
also $\left|\sin\phi\right|\approx 0$ we find
\begin{equation}
  \label{eq:18}
  (1.0\pm0.7)\% = (-2.5^{+1.4}_{-1.6})\% \cos \delta - (0.0\pm
  1.5)\% \sin \delta \ ,
\end{equation}
which implies a certain distribution for $\cos\delta$. Due to the sign
difference between $y_{CP}$ and $y'$ and due to the relative smallness
of $x'$ it is expected that this distribution of $\cos\delta$ will be
biased to negative values. By a full analysis, considering the
measured values of $A_m$ and $\sin\phi$ we can characterize the bias
by stating the total confidence level value:
\begin{equation}
  \label{eq:cosdelta}
  \cos\delta \lesssim 0.7 \qquad (95\% \text{ C.L.}) \; ,
\end{equation}
(and $\cos\delta \lesssim 0.0$ at $68\%$ C.L.).

Since we have now a distribution for $x'$, $y'$ and $\cos \delta$, we
may invert \eqref{eq:xpyp} to solve for $x$ and $y$:
\begin{eqnarray}
  \label{eq:xy}
  x& = & x' \cos \delta - y' \sin \delta \ , \\
  y& = & y' \cos \delta + x' \sin \delta \ . \nonumber
\end{eqnarray}
We note that the signs of $x$ and $y$ in~\eqref{eq:xy} are not
measured by current experimental results. Since the measured value for
$x'$ is distributed around zero the sign for $y$ is determined by the
sign of $y'$ which, in turn, depends on the sign of $\cos \phi$. This
sign is not provided by any measurement (all we know is that
$\left|\cos\phi\right| \approx 1$). Similarly, the sign of
$x$ is determined by the sign of both $y'$ and $\sin\delta$, which are
not measured.

The resulting distributions for $x$ and $y$ are therefore in the form
of two superimposed distributions for the two possible sign choices
(denoted by the $\pm$ sign). We obtain:
\begin{eqnarray}
  \label{eq:19}
  x & \approx & (\pm 2.8 \pm 2.5)\% \ , \\
  y & \approx & (\pm 0.9 \pm 3.6)\% \ . \nonumber
\end{eqnarray}

We note that these values are different from those quoted
in~\cite{Groom:2000in} where it is assumed that $\delta=\phi=0$. When
we consider the distribution of $\cos\delta$, the bound on
$x$ (and hence the bound on $\Delta m_D$) is weakened by a factor of
about $2.2$. The bound on $y$, however, (and hence the bound on
$\Delta\Gamma$) is strengthened.  For comparison,
table~\ref{tab:compare_xandy} shows the $95\%$ C.L.\ ranges for $x$ and $y$
in the two cases: One which assumes $\cos\delta=1$ and $\cos\phi=1$,
and one which takes the values mentioned.
\begin{table}
  \caption{Comparison between mass and width difference parameters at
    $95\%$~C.L. with different assumption on mixing parameters.}
  \label{tab:compare_xandy}
  \begin{ruledtabular}
    \begin{tabular}{cc}
      Assuming $\cos\delta=1$,$\cos\phi=1$ & No assumption\\
      \hline
      $\left|x\right| \lesssim 2.9\%$ & $\left|x\right| \lesssim
      6.3\%$ \\
      $-5.8\% \lesssim y \lesssim 1.0\%$ & $\left|y\right| \lesssim
      4.6\%$ \\
    \end{tabular}
  \end{ruledtabular}
\end{table}

We evaluate now the \DDbar mixing amplitude.
Taking the average decay width~\cite{Groom:2000in}
\begin{equation}
  \label{eq:20}
  \Gamma_D = (1.595 \pm 0.011)\times 10^{-9}\; \text{MeV}\ ,
\end{equation}
and using \eqref{eq:16}, we obtain a distribution for $M_{12}$ which is
maximal near zero:
\begin{equation}
  \label{eq:21}
  \left|M^D_{12}\right| \leq 
    6.2\times 10^{-11}\; \text{MeV }\qquad (95\%\ \text{C.L.})\;\;,
\end{equation}
(and $  \left|M^D_{12}\right| \leq 3.3\times 10^{-11}\; \text{MeV }$
at $68\%$ C.L.).

It is interesting to compare this value to the ones obtained by using
some simplifying assumptions. First, assuming no CP violation in
mixing, we set $A_m=0$ but allow for $\delta,\phi \neq 0$. We get
\begin{equation}
  \label{eq:22}
  \left|M^D_{12}\right| \leq 5.4\times 10^{-11}
  \text{MeV } \qquad (95\%\ \text{C.L.})\;.
\end{equation}
Second, we set $A_m=\phi=0$ and allow $\delta\neq 0$. We get
\begin{equation}
  \label{eq:23}
    \left|M^D_{12}\right| \leq 4.0\times 10^{-11}
  \text{MeV } \qquad (95\%\ \text{C.L.})\;.
\end{equation}
Third, we set $\delta=0$, but allow $A_m,\phi\neq 0$. We get
\begin{equation}
  \label{eq:25}
    \left|M^D_{12}\right| \leq 3.9\times 10^{-11}
  \text{MeV } \qquad (95\%\ \text{C.L.})\;.
\end{equation}
Last, we set $A_m=\phi=\delta=0$ and get\footnote{Actually, it is
  enough to assume $A_m=\delta=0$ since, in this case, the value of
  $\phi$ affects only $y$, which does not contribute to $M^D_{12}$.}
\begin{equation}
  \label{eq:26}
  \left|M^D_{12}\right| \leq 2.3\times 10^{-11}
  \text{MeV } \qquad (95\%\ \text{C.L.})\;.
\end{equation}
This is the value which appears in~\cite{Groom:2000in}.
Thus, allowing CP violation, the resulting constraint is about $2.7$
weaker (i.e.\ larger) then the one which is obtained with the maximal
set of assumptions.

\section{Conclusions.}
\label{sec:conclusions}

We interpret the most recent data from the experimental searches for
\DDbar mixing. Allowing CP violation in mixing, we obtain the upper
bound
\begin{equation}
  \label{eq:24}
    \left|M^D_{12}\right| \leq 6.2\times 10^{-11}\; \text{MeV
    }\qquad (95\%\ \text{C.L.})\;,
\end{equation}
which is $2.7$ times weaker than the naive calculation. 

The actual upper bound for \DDbar mixing amplitude depends, therefore,
on the model in question. Assuming that CP is conserved in \DDbar
mixing, as is the case in the standard model, the bound is the one
in~\eqref{eq:23}. (If, in addition, one is willing to assume
  that $SU(3)$-flavour symmetry holds in $D$ decays, the bound is
  given by~\eqref{eq:26}.) For a more general model, with new sources of CP
violation, eq.~\eqref{eq:24} gives the present bound. Taking into
account this weaker bound leads to modifications \cite{alignment}
compared to analyses that consider only the CP conserving
bound~\cite{Hou:2001ua,Chang:2001ah,Chua:2001dd}.

\begin{acknowledgments}
  I thank Yossi Nir for his help and guidance.
\end{acknowledgments}



\end{document}